# Water Fibers

Mark L. Douvidzon[1], Shai Maayani[2], Leopoldo L. Martin[2] and Tal Carmon[2]

Technion, 1 Department of Nanoscience and Nanotechnology RBNI and 2 Department of Mechanical Engineering.

**Fibers constitute the backbone of modern communication[1] and are used in laser surgeries[6]; fibers also genarate coherent X-ray[2,3], guided-sound[4] and supercontinuum[5]. In contrast, fibers for capillary oscillations, which are unique to liquids, were rarely considered in optofluidics[7–9]. Here we fabricate fibers by water bridging[10–17] an optical tapered-coupler[18] to a μlensed coupler[19]. Our water fibers are held in air and their length can be longer than a millimeter. These hybrid fibers co-confine two important oscillations in nature: capillary- and electromagnetic-. We optically record vibrations in the water fiber, including an audio-rate fundamental and its 3 overtones in a harmonic series, that one can hear in soundtracks attached. Transforming Micro-Electro-Mechanical-Systems [MEMS] to Micro-Electro-Capillary-Systems [MECS], boosts the device softness by a million to accordingly improve its response to minute forces. Furthermore, MECS are compatible with water, which is a most important liquid in our world.**

At a scale less than the thickness of a hair (Latin: capillus), cohesive forces between intimate particles of liquid result in tension that pushes to minimize the liquid surface. In contrast with light and sound, which are common to all phases of matter, capillary waves are unique to liquids. Capillary oscillations were widely studied in flat interfaces[20–22] and in droplets[23,24], which makes it natural to extend such opto-capillary effects to fibers. For fabricating an opto-capillary fiber, one can take a liquid-core fiber[25] and remove its solid cladding so that the liquid interface can move instead of being bounded between solid walls. But fabricating a suspended water string raises some structural hurdles. For mitigating these structural challenges and building a floating water-bridge, we use a practice known since 1893. Back then, an electric potential-difference between two beakers resulted in an unsupported water-thread that emerged while fluidically connecting the two reservoirs[10]. Such water bridges were demonstrated at length longer than 2.5 centimeter[14] or with a diameter thinner than 20 nano-meters[12]. Yet, and unlike water fountains, water bridges were never used for guiding light. These types of water threads can, in principle, guide electromagnetic-, acoustical- and capillary-modes, which make them attractive for multi-sensory labs-on-chips[7–9], where several different waves can participate. This is in contrast to multi-spectral systems were different frequencies of the same wave participate. In constructing such water bridges, one harnesses dielectric- and interfacial-tensions to serve in carrying the bridge's weight[15–17]. With close descriptions encounter in literature for the ratio between the dielectric- and surface-tensions; we consider both of these tensions in carrying the bridge weight, and get the ratio between them from our experiment as will be explained in what follows.

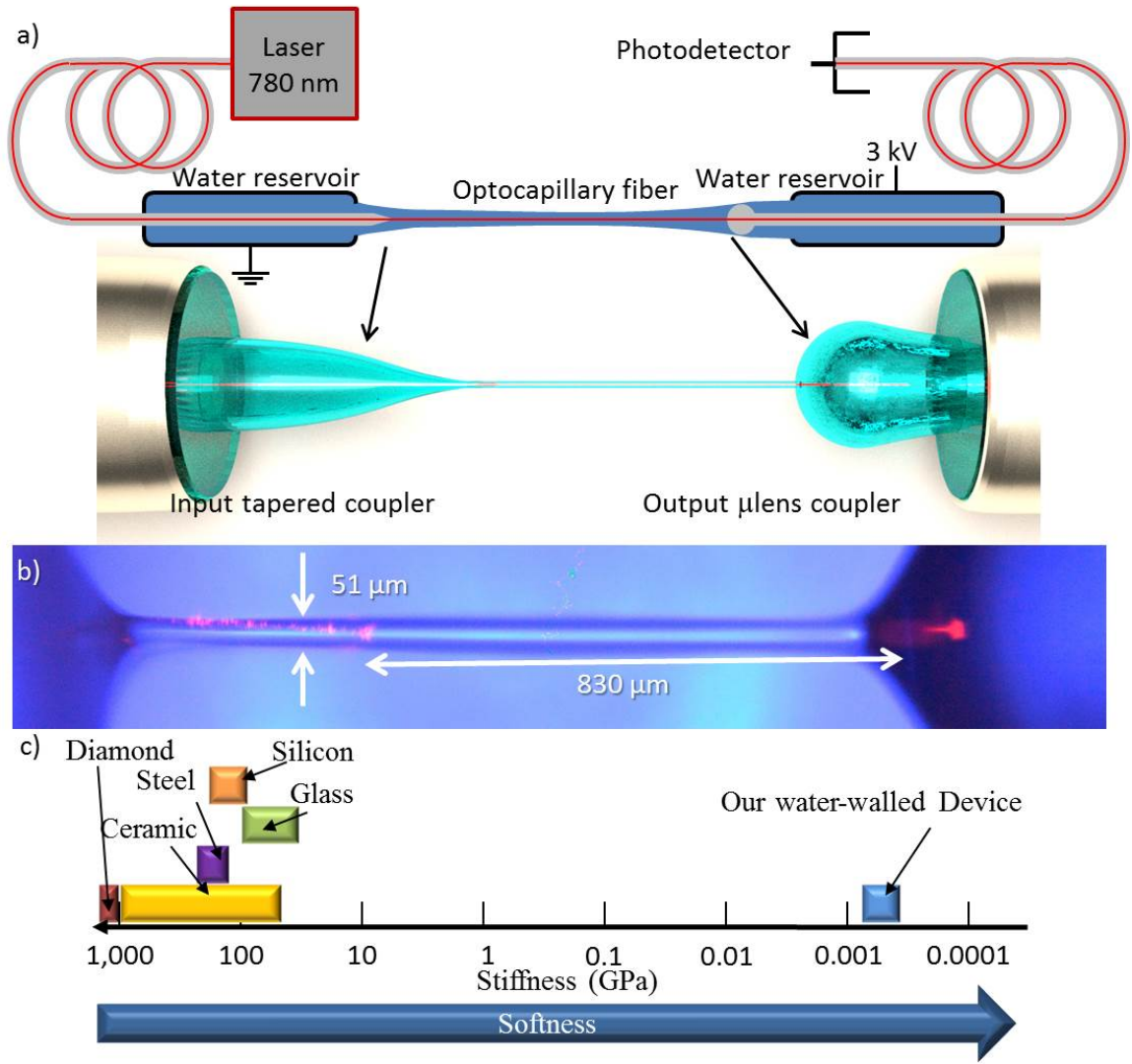

**Figure 1 Fabricating a water fiber:** an Illustration **(a)** and a Photograph **(b)**. **(c)** The softness of our water-walled waveguides in comparison with common solids. The length of the fiber is measured from the tip of the tapered fiber to the beginning of the spherical lens (solid to solid). The fiber was back illuminate at blue in order to improve resolution while providing contrast in respect with the red light that is propagating in the fiber. Residual scatterings here (red) are from the solid tapered-fiber and lensed-fiber and not from the liquid.

*Fabricating* an opto-capillary fiber starts from its bases that we make by inserting silica fibers into water-containing micropipettes (Fig. 1a). The micropipettes provide water from a practically unlimited aquatic reservoir, to replenish the water bridge and compensate for evaporation. In regard with this pipette, it is important that its internal side will be hydrophilic while its outer side is hydrophobic[26], since otherwise water will leak and climb on the outer side of the pipette[27]. We fabricate the input coupler by tapering a silica fiber[18]. Adiabatically tapering this glass fiber, is done to a slope that is shallower than ~1/20, which is preferred here as it can assist in transmitting more than 99% of the light[28]. We then fabricate the output coupler by reflowing the end of a silica fiber to the form of a lens that focuses the incoming light into the fiber core[19]. More details on the fabrication procedure can be found in the method section. We now position the two sides of the bridge so that they are almost touching each other. A minute fluidic contact is then spontaneously established between the two sides. We then raise the electrical potential difference between the two optical couplers while increasing the distance between them. As one can see in Figure 1, a water bridge is then formed between the optical couplers while light is inserted from one side and coupled out to a photodiode through the other side.

We now *characterize* the water fiber by coupling in a 770nm laser light from its tapered-fiber coupler and measure an optical transmission larger than 54% through a 0.83 millimeter long water fiber (Fig.1b).

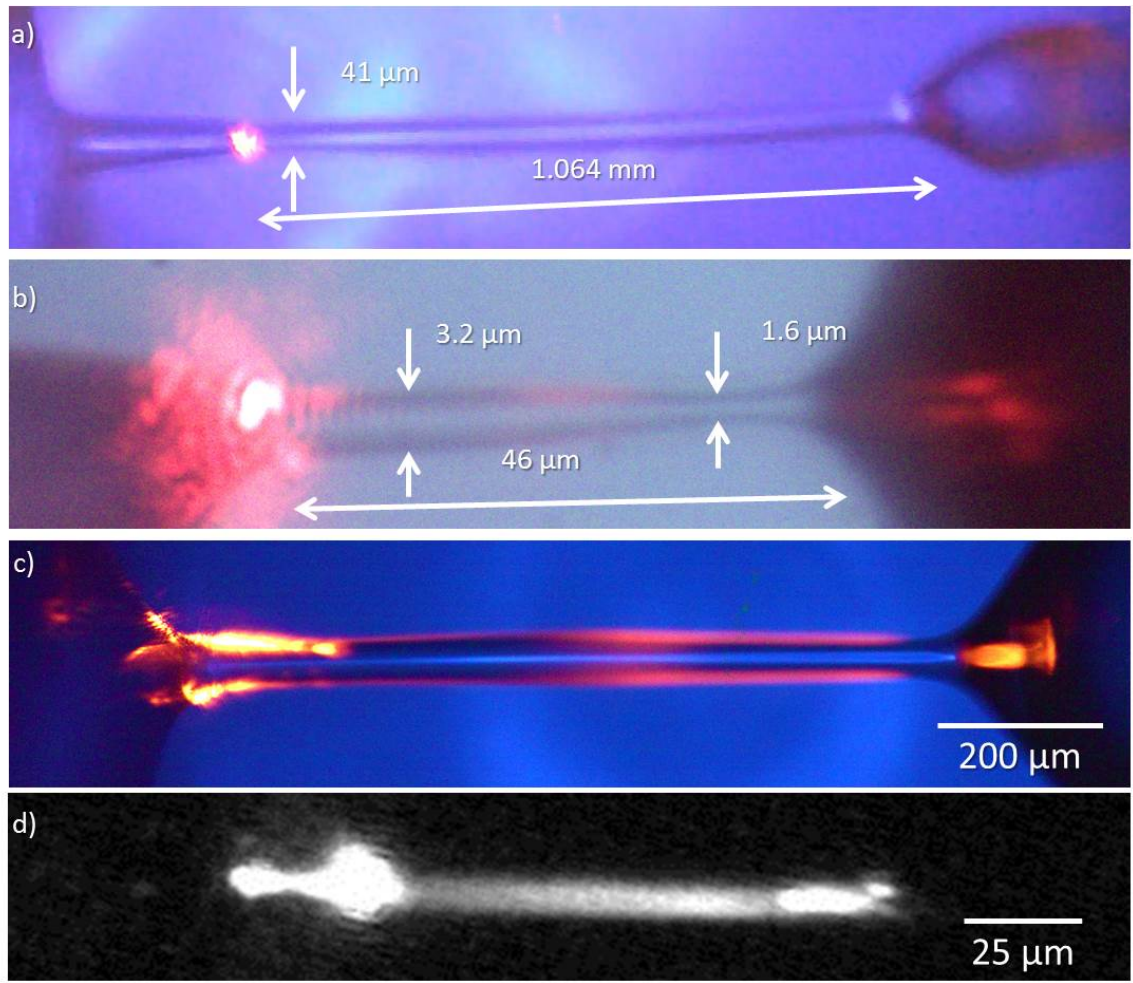

**Figure 2 Characterizing a water fiber.** (**a**) A longer than a millimeter water fiber. (**b**) a thin water fiber, (**c**) surface scattering from the water fiber followed by a florescent measurement (**d**) indicating that light propagates inside the fiber.

As for the water-fiber length, our optocapillary fibers could extend over 1.06 millimeter length (Fig. 2a). The radius of the fiber could be smaller than 1.6 µm (at L=46 µm, Fig. 2b); and at some regions, as small as 800 nm (Fig. 2b). The water fibers were maintained as long as necessary for our experiments, which were typically taking 10 minutes. As one can see in Figure 2c, light scattering seems to prefer the fiber outer surface. Such scatterings were extensively studies in flat interfaces[20–22] and are therefore not the focus of this research. Yet, we had to verify here that light indeed propagates in the water fiber and not just near its surface as one might think from looking at Figure 2c. In order to exclude this possibility that light prefers to propagate near the fiber interface, we were using a fluorescent dye to map light propagation. The fluorescent emission from the places where light propagated (Fig. 2d), instead of from places where light is scattered (Fig. 2c), confirms that light indeed propagates inside the fiber. Some scatterings seen in Figure 2c are therefore the result of thermal capillary waves that reside at the water-air interface and make the surface rough at a sub Angstrom scale[20–24].

In our *experiments,* we will optically interrogate the capillary resonances of the water fiber via its effect on optical transmission, which is measured as shown in figure 1a. As opposed to surface capillary waves that travel near the water-phase boundary (Fig. 2c and[20–24]), we will show here body waves that move through the interior of the water structure. In more details and similar to a vibrating guitar-string, the water thread can vibrate at frequencies

$$(1) \quad f_i = \frac{i}{2L}\sqrt{\frac{T}{\rho`}},$$

where $\rho`$ is the string mass density per unit length, $L$; $T$ is the water-bridge tension and $i=1,2,3\ldots$ is the eigenindex describing the string length in units of half capillary-

wavelength. The water-fiber tension is affected by dielectric-[15] and surface- tensions [16] and to be[17]

$$(2)\ T = a \cdot (\frac{\pi D^2}{4})[\frac{1}{2}\varepsilon_0(\varepsilon-1)E^2] + b \cdot \gamma \pi D$$

where $D$ is the bridge diameter; $\varepsilon_0$ is the vacuum permittivity; $\varepsilon$ is the relative permittivity of the liquid; $\gamma$ the surface tension and $E$ is the electric field along the bridge. $a$ and $b$ here will be fitted experimentally to describe the ratio between electrostrictive-forces[15,17] and interfacial-tension[16,17] that are co-carrying the water-bridge weight.

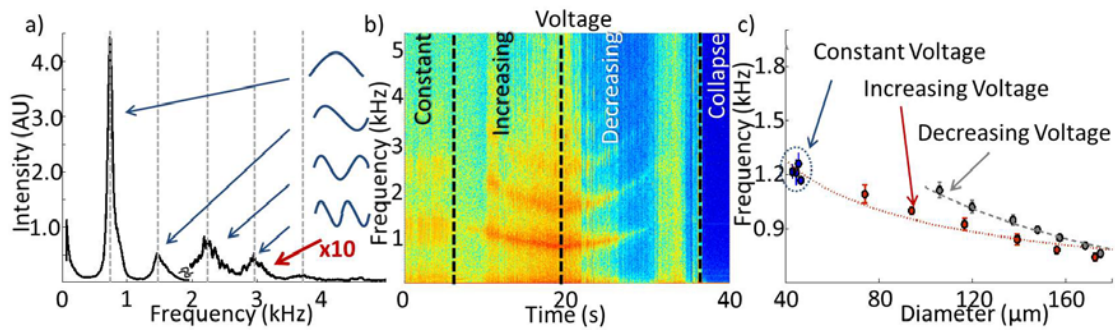

**Figure 3 Experimentally measuring the "guitar-string" modes of the water fiber. (a)** Fluctuation spectrum reveals a fundamental mode of vibration together with its 3 overtones at integer multiplications of the fundamental frequency (dash lines). **(b)** Fluctuation spectrogram of a 0.94 mm long fiber along 40 seconds, in which the voltage first stays constant, then increased and then decreased which changes the fiber diameter. **(c)** The fundamental frequency of the fiber as a function of the fiber diameter (circles) together with what calculated with equations 1-2 (lines) where the ratio between dielectric and surface forces (a/b) is a free parameter. **(a-c)** fiber length is 0.94 mm and oscillation is optically interrogated with a photodetector (Fig. 1).

As one can see in Figure 3a, we are experimentally measuring spectral peaks at the optical transmission through the water fiber. In more details, a strong spectral peak, that we call fundamental, is seen at 744 Hz, with higher harmonics at integer multiplications of this frequency (Fig. 3a, dashed lines). Assuming that these oscillations originate from the string like resonances of the water-finer, we would expect that increasing the diameter of the fiber (and its mass) would reduce their frequency as predicted by equations 1-2. To confirm this assumption we continuously change the diameter of the water fiber by changing the voltage (Fig. 3b), which indeed results a chirp in the resonance frequencies as predicted by equation 1-2 (Fig. 3c). Hysteresis was evident in the frequency with maximal value of 20%.

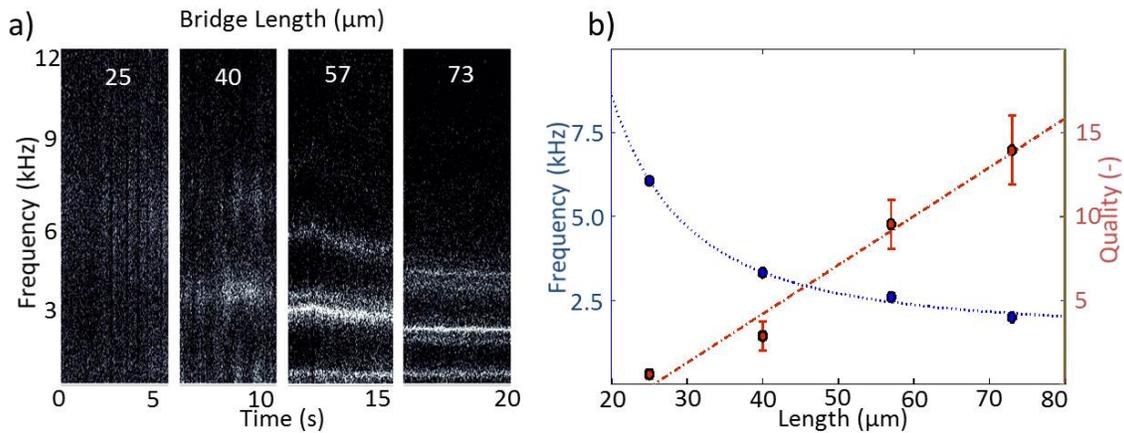

**Figure 4 Mechanical Eigenfrequencies of the water fiber. a)** Spectrogram of a liquid fiber with diameter of 5.5 µm and varying length. **b)** Eigenfrequency and quality factor as a function of the fiber length. The blue line is a fit to equation 1 where *a* and *b* are the free parameter. The red line is a guide for the eye.

In a different experiment, we are monitoring the oscillation frequency while changing the length of the water fiber (Fig. 4a) and keeping its diameter as constant as possible

(Methods table a). Since tension, *T*, is a function of the electric field *E*, which declines with fiber length *L*, we expect the capillary eigenfrequency to scale inversely with length. As expected, the oscillation frequency is indeed measured to decrease with length (Fig. 4b, blue circles) as predicted by equations 1-2 (Fig. 4b, blue line). As for the linewidth of the resonances that we measure, they narrow with length. The energy loss in this system relates to the viscosity of water, which is more significant at higher rates. For this reason, linewidth narrowing at longer fibers (with low rates) is expected (Fig. 4b). In more details, the capillary quality factor, Q, for our fiber is a dimensionless parameter that is inversely proportional to energy losses and linewidth. Q is defined by the energy it stores in units of loss during one cycle of oscillation and is calculated to be[29]

$$(3) \quad Q = 1/\alpha \Lambda, \quad \alpha = \frac{8\mu\pi^2}{\rho v_g \Lambda^2} \quad \text{and} \quad v_g = \sqrt{\frac{4.5\pi\gamma}{\rho\Lambda}},$$

where $\alpha$ is the absorption coefficient for capillary waves, $\Lambda$ is the capillary wavelength, $\mu$ is the liquid viscosity, $v_g$ is the group velocity for capillary waves and $\rho$ the liquid mass density. As expected, *Q* decreases with length (Fig. 4, red circles), though not with the exact nonlinearity predicted by Eq. 3, probably because our thin-string approximation breaks for the shortest opto-capillary fiber. The highest capillary quality factor that we could measure was 14 at 1991 Hz (Fig. 4b).

The capillary resonances that we optically record here are at the audio band, meaning that we can listen to them without any frequency conversion. We therefore provide soundtracks where one can hear the evolution of the resonance from nearly white noise for a short (25 µm) water fiber to a distinct tune at a defined frequency for the long (73 µm) water fiber (complimentary a). As for the continuously changing water fiber diameter (complimentary b), we can hear the chirp toward lower (higher) frequency when the water fiber diameter increases (decreases). Interestingly, the sound that the water fiber makes in this experiment (Fig. 3b) resembles the sound of a passing jet-plane and its related Doppler shift.

To conclude, similar to optical-waveguides, which are significant for enhancement of electromagnetic phenomena by canceling-out diffraction, capillary fibers can now confine interfacial-tension related phenomena. As our fibers confines both light and capillary waves, they permit optical interrogation of the capillary oscillations. Our water-fiber can therefore extend the current state of the art in nonlinear optics[30] to enable also nonlinear opto-capillary interactions as well as lab-on-chip applications where most of the external walls are made of water and capillary phenomena are confined, resonantly enhanced and optically interrogated.

**Methods**

The input tapered-fiber coupler is Corning HI 780 C, tapered over a hydrogen flame as explained in[18]. The output silica lensed-fiber coupler is Thorlabs FTO 30 that was reflowed using a CO2 laser as explained in[31].

The micropipettes are Round Borosilicate Glass Capillaries in Convenience Vials.

We use deionized water with a resistivity of 18 MΩ·cm.

The fluorescent dye (Fig.2d only) is ADS, near infrared Absorption dye, ADS 780WS and its near IR fluorescent emission, at 810 nm, was photographed through a long-pass optical filter that blocks the visible.

Other parameters are as appear in the following table:

|  | Water fiber | | | Pipette's internal diameter | |
|---|---|---|---|---|---|
|  | Length [μm] | Radius [μm] | Potential [V] | Taper side [μm] | Lens side [μm] |
| Fig. 1b | 830 | 51 | 6000 | 850 | 850 |
| Fig. 2a | 1064 | 20.5 | 6000 | 850 | 850 |
| Fig. 2b | 46 | 1.6 - 0.8 | 1500 | 150 | 850 |
| Fig. 2c | 820 | 32.5 | 5000 | 850 | 850 |
| Fig. 2d | 110 | 4.75 | 3000 | 150 | 150 |
| Fig. 3 | 940 | 20 - 90 | 3000 - 8000 | 850 | 850 |
| Fig. 4 | 24 - 73 | 2.7 - 3 | 2500 | 150 | 850 |

**Supplementary Information**

Complementary a.avi – Varying the length video and sound file

Complementary b.avi – Varying the diameter video and sound file

**Acknowledgements**

Funding. This research was supported by ICore: the Israeli Excellence center 'Circle of Light' grant no. 1902/12 and by the Israeli Science Foundation grant no. 2013/15

**Author Contributions**

L.L.M. fabricated device. M.L.D. performed the experiments and developed the theory. S.M. analyzed the data. T.C. supervised the research.

**Correspondence** and requests for materials should be addressed to

Tal.Carmon@gmail.com